\documentclass[conference,letterpaper]{IEEEtran}
\usepackage{graphicx} 
\usepackage{color} 
\usepackage{ulem} 
\usepackage{url}
\graphicspath{{images/}}
\usepackage[cmex10]{amsmath}
\usepackage{amssymb}

\begin{document}

\title{CLIFT: a Cross-Layer InFormation Tool to perform cross-layer analysis based on real physical traces}  

\author{
\IEEEauthorblockN{Nicolas Kuhn$^{1,2}$, Emmanuel Lochin$^{1}$, J\'{e}r\^{o}me Lacan$^{1}$, Roksana Boreli$^2$, Caroline Bes$^3$, Laurence Clarac$^3$}
\IEEEauthorblockA{\\
$^1$Universit\'{e} de Toulouse, ISAE, TeSA, Toulouse, France\\
$^2$NICTA Sydney, Australie\\
$^3$Centre National d'Etudes Spatiales, CNES\\
}
}

\maketitle

\begin{abstract}
Considering real physical (MAC/PHY) traces inside network simulations is a complex task that might lead to complex yet approximated models.
However, realistic cross-layer analysis with the upper layer and in particular the transport layer cannot be driven without considering the MAC/PHY level.
In this paper, we propose to cope with this problem by introducing a software that translates real physical events from a given trace in order to be used inside a network simulator such as $ns$-2.
The main objective is to accurately perform analysis of the impact of link layer reliability schemes (obtained by the use of real physical traces) on transport layer performance. 
We detail the internal mechanisms and the benefits of this software with a focus on 4G satellite communications scenarios and present the resulting metrics provided by CLIFT to perform consistent cross-layer analysis.
\end{abstract}

\section{Introduction}
\label{sec::introduction}

The increase of wireless and satellite links in current networks introduce challenging issues. In the case of Land-Mobile Satellite (LMS) channels, the most powerful codes cannot recover lost data, due to long bit-errors bursts at the physical layer~\cite{lms_state}. 
To mitigate these losses, simple retransmission (if a return channel is available) or the use of erasure codes with or without retransmission of a repair Link Layer Data Unit (LLDU) are often enabled at the link layer. In~\cite{link_layer_rel}, the authors lead an extensive study on the reliability schemes that can be implemented at the link layer level: Forward Error Coding (FEC), Automatic ReQuest (ARQ), Selective-Repeat Automatic ReQuest (SR-ARQ) and Hybrid-Automatic ReQuest type II (HARQ-II). Introducing reliability schemes at this level can prevent the transport layer from decreasing its congestion window in case of isolated errors. In the context of high latency links, these techniques might introduce critical delays and impact on the transport layer protocols performance~\cite{high_bd_del_prod}. 

As a result, specific transport protocols have also been designed to overcome these problems (\textit{e.g.} TCP Hybla for satellite links, TCP Westwood for wireless links, CUBIC for high delay bandwidth product networks). Nevertheless, errors encountered by a LMS channel might greatly degrade their performance. In this context, it is essential to study the impact of the most recent codes (at both physical and link layers) on the transport protocols performance. Unfortunately, and to the best of our knowledge, there is no tool allowing to easily perform such study.  

Preliminary studies have explored TCP performance over link layer ARQ protocols in wireless environment \cite{tcp_arq_1,tcp_arq_2} and in the context of 4G satellite system downlink~\cite{tcp_4G}. One recent proposal~\cite{aimd} has developed analytical tools in order to evaluate the impact of reliability schemes at the link layer on transport layer protocols while some others \cite{ll_ns2,barakat_fec_arq} attempt to consider link-layer data unit.
Nowadays, there is a clear lack of an exhaustive tool allowing to evaluate currently deployed protocols (CUBIC in GNU/Linux or Android and TCP Compound in Windows operating systems) over realistic MAC/PHY layer traces.

Our proposal, called Cross-Layer InFormation Tool (CLIFT), links an updated and maintained network simulator, $ns$-2, with recent lower layers codes performing over real physical channel state traces. CLIFT is not a physical layer simulator (opposed to \cite{ns-3}) but a way to take into account physical layer traces inside a network simulator. Therefore, CLIFT allows to study the impact of link layer reliability schemes, as a function of a given physical channel, on transport protocols performance. The rationale of such approach is to play MAC/PHY traces (CLIFT allows to read several existing traces format) either empirically measured or generated by a physical layer emulator or simulator.


The rest of the paper is organised as follows: in Section~\ref{sec::arch_soft}, we briefly detail the structure of our tool. In Section~\ref{sec::link_layer}, we present the physical layer traces and how CLIFT can consider link layer reliability schemes. Then, we detail the problems encountered in the development of the queuing module for $ns$-2 in Section~\ref{sec::ns2}. We illustrate the potential of our tool through an example in Section~\ref{sec::illustration}. 
Finally, we conclude in Section~\ref{sec::conclusion}.

\begin{figure*}[!htb]
    \begin{center}
	\includegraphics[width=0.9\linewidth]{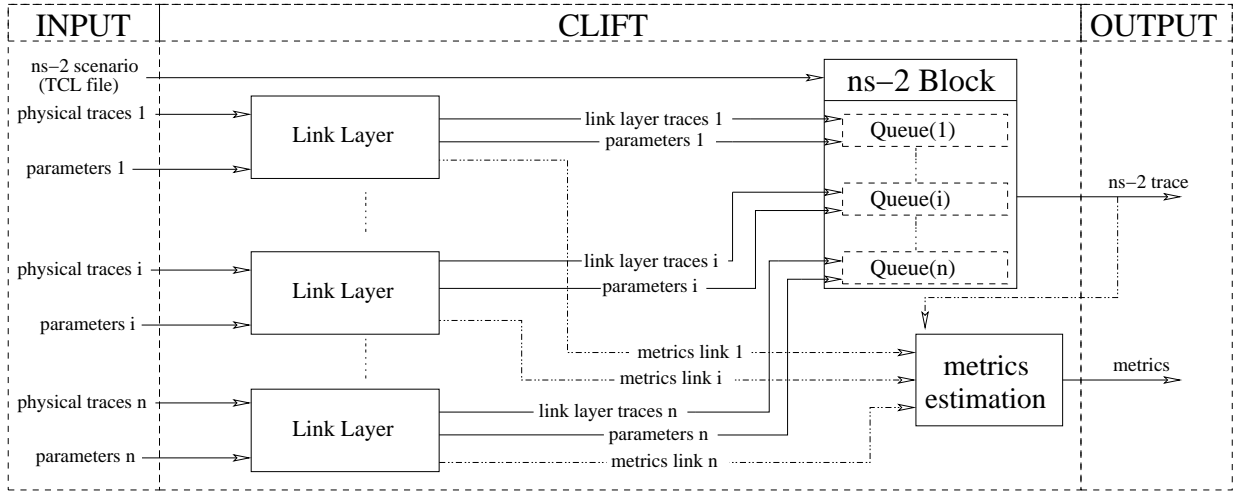}
	\caption{Structure of software}
	\label{fig::internal_struc}
    \end{center}
\end{figure*}

\section{Software Architecture}
\label{sec::arch_soft}

Before diving into the software details, we propose in this section to firstly present the overall structure of CLIFT and the linkages between each internal component. We also define how to define a simulation and present the metrics provided by CLIFT. 

\subsection{CLIFT main internal components}
\label{sec::arch_soft_achievements}

CLIFT is based on three main components (cf. Figure~\ref{fig::internal_struc}): 

\begin{itemize}
\item the \textit{link-layer} component: for each link of the network, CLIFT loads a given physical trace. We explain in Section~\ref{sec::link_layer} how reliability schemes at this layer can be taken into account;
\item the \textit{$ns$-2 block} component: we developed a queuing module in $ns$-2 that loads these link layer trace to schedule the transmission of the transport layer packets. The $ns$-2 module implementation is detailed in Section~\ref{sec::ns2};
\item the \textit{metric evaluation block} component: that provides the resulting measures.
\end{itemize}

\subsection{Defining a complete simulation}
\label{subsec::arch_soft_complete_simu}
A simulation is performed following the $ns$-2 standard procedure where the user needs: 
\begin{itemize}
\item to define the network structure through a standard TCL $ns$-2 simulation file;
\item for each link, to define a parameter file and provide a physical layer trace;
\item then to run $ns$ simulation.
\end{itemize}
For each link, CLIFT adapts the measurements trace depending on the possible reliability schemes introduced and analyses the traces
to compute the relevant metrics.

\subsection{Metrics evaluation}
\label{subsec::arch_soft_metrics}

Two kinds of metrics are returned by CLIFT:

\begin{itemize} 
\item link layer level metrics: throughput efficiency, delay, retransmission distribution, erasure distribution;
\item transport layer level metrics: used resources (percentage of the bandwidth), delay, number of RTO events, retransmission  distribution, throughput,  queuing delays.
\end{itemize}

All these metrics allow to perform a consistent cross-layer analysis. This will be later illustrated in Section~\ref{sec::illustration}.

\section{Dealing with physical layer trace}
\label{sec::link_layer}

One of the main advantage of CLIFT is to bring real physical traces into network simulator. In this section, we thus focus on the physical layer trace format and present the erasure codes that can be optionally applied to these traces.

\subsection{Physical layer trace format}
\label{subsec::phys_trace}

\begin{figure}[!ht]
    \begin{center}
	\includegraphics[width=1\linewidth]{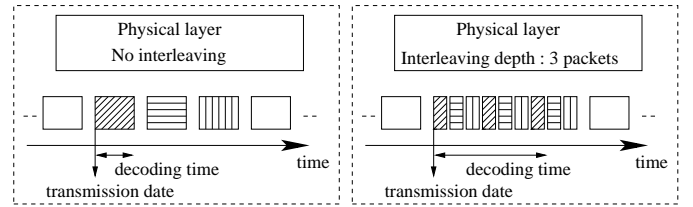}
	\caption{Physical layer traces: transmission and decoding times}
	\label{fig::phys_trace_layer}
    \end{center}
\end{figure}

CLIFT accepts, as an input, several physical traces format: both measured (as those provided in CRAWDAD\footnote{http://crawdad.cs.dartmouth.edu}) or generated by a physical layer emulator \cite{judd04} or a simulator \cite{ns-3}. As an example, we propose the use of OFDM and TDM simulators from CNES\footnote{CNES is a government agency responsible for shaping and implementing France's space policy in Europe, see \url{http://www.cnes.fr/}.} that takes into account realistic satellite links characteristics, such as satellite orbits or recent correcting codes to generate physical layer traces~\cite{use_ofdm}. Each packet sent at the physical-layer level is characterised by an transmission date and a decoding time. In Figure~\ref{fig::phys_trace_layer}, in order to better assess the link between transmission date and decoding time, we illustrate how they are affected by interleaving at the physical layer. The transmission date is linked to the bandwidth and the length of the code at the physical layer. The decoding time is linked to the duration of the interleaving, the channel state and the transmission time. As CLIFT can load any physical layer trace compliant with this format, they can be either real measured traces or traces obtained by a physical layer simulator. Therefore, the main achievement of CLIFT is that real measured channel evolutions can be considered, while modeling such channels might lead to approximation and errors.

The decoding time is composed of the different delays introduced by the reliability schemes at the physical layer level (interleaving and recovery delay). We denote: 
\begin{itemize}
\item LLDU as one Link Layer Data Unit;
\item $td_i$ as the transmission date of $LLDU_i$;
\item $dt_i$ as the decoding time of $LLDU_i$;
\item $dt_i=0$ as the erasure event of $LLDU_i$.
\end{itemize} 
At $t=\frac{RTT}{2}+td_i+dt_i$, the physical-layer delivers $LLDU_i$ to the link layer, if there is no supplementary delay (congestion, queuing, ...). 

\subsection{Link layer Model}
\label{subsec::link_layer_schemes}

The traces considered by CLIFT can be MAC/PHY traces that implement or not reliability schemes. If we use traces that do not enable reliability mechanisms at the MAC level (\textit{e.g.} ARQ or H-ARQ), we could also perform a pre-treatment over this traces with a tool such as TMT~\cite{link_soft_tmt}, PPR~\cite{link_soft_ppr} or DUMMYNET~NETEM\cite{link_soft_dummynet} that allow to apply reliability mechanisms up the MAC level. Basically, these tools allow to modify the PHY traces, following a given reliability mechanism used at the MAC level, by recomputing the transmission slots.
The principle is as follows : the decoding time of one erased LLDU is linked to the reliability scheme involved to estimate the time when the recovered LLDU must be sent. The supplementary time introduced by the link layer reliability scheme, denoted $dti'$, is the time needed to obtain ($td_R$) and decode ($dt_R$) the $LLDU$ that enables the recovery of $LLDU_i$: $dti'=td_R+dt_R-td_i$. A physical layer data unit will be delivered to the link layer at $t=\frac{RTT}{2}+td_i+dt_i'$.
 
We detail below commonly used reliability schemes:
\begin{itemize}
\item FEC: The sender sends $N_D$ data and $N_R$ repair LLDU. The link layer can repair a maximum number of $N_R$ LLDU;
\item SR-ARQ: The link layer retransmits the lost LLDU;
\item HARQ of type II: This mechanism is a combination of FEC and SR-ARQ. After a first transmission of a FEC block, including data and repair LLDU, HARQ-II allows the sender to send additional repair LLDU when a recovery is not possible at the receiver side.
\end{itemize}
We denoted HARQ\,($N_D$,$N_D+N_R$) (or FEC\,($N_D$,$N_D+N_R$)), where $N_D$ is the number of data LLDU and $N_R$ the number of repair LLDU.

As a result, one other main achievement of CLIFT is to consider the most recent link layer reliability schemes applied on realistic physical layer traces.

\section{Internal software principle}
\label{sec::ns2}

We schedule the transmission of the IP packets depending on the link layer traces (section~\ref{subsec::link_layer_schemes}): we introduce a new queuing module in $ns$-2 that loads these traces and determines when a packet can be recovered by an upper layer (depending on the reliability schemes introduced) and sent. The queuing system in $ns$-2 is mainly driven by the following entities: packets (with arrival times and services times attributes) and queue (with empty and non-empty attributes)~\cite{ns2_manual}.

The function \texttt{enqueue()} is called when a packet arrives in the queue. When the channel is idle, the function \texttt{dequeue()} is called to transmit the packet chosen depending on the queuing mechanism. We aim to modify these functions following the scheduling read in the link layer trace.

\subsection{Add an IP packet in the queue: the \texttt{enqueue()} function}
\label{subsec::n2_enque}

One IP packet is divided into $m$ LLDU ($LLDU_n, ... LLDU_{n+m}$), where $m=E ( {\frac{size(IP packet)}{size(LLDU)}} ) +1$. When a $IP packet_i$ is enqueued at $Te_i$, we look in the link layer trace for the LLDU that matches $td_n \leqslant Te_i<td_{n+1}$. Over the $m$ LLDUs, we compute $Td_i=\max_{k \in [n, n+m]}(td_k+dt_k)-Et_i$, where $Et_i$ is the transmission time of $IP packet_i$ and $Td_i$ the transmission date of $IP packet_i$. Indeed, $\max_{k \in [n, n+m]} (td_k+dt_k)+RTT/2$ represents the date when $IP packet_i$ is delivered to the receiver. 

We handle the case $Td_i<Te$ since $ns$-2 is a event-driven simulator: for example, this event might occur when erasure codes are used, and bursts of LLDUs are forwarded to the upper layer. With a FEC code, if LLDUs are lost, they are all rebuilt at the same time with the reception of the $N_R^{th}$ LLDU. 

\subsection{Remove an IP packet from the queue: the \texttt{dequeue()} function}
\label{subsec::n2_deque}

As soon as an IP packet enters the queue, we introduce a timer which value is set depending on the transmission date of the LLDU packets the IP packet is broken down into. When the timer expires, the method \texttt{dequeue()} is called and the corresponding IP packet is sent. We seek the next IP packet to be sent and we set the timer to its new value. 
We adapt the timer value if:  
\begin{itemize}
\item an IP packet is enqueued and there is no other packet in the queue;
\item an IP packet is enqueued and its transmission date is earlier than those of the packets in the queue;
\item an IP packet has to be removed from the queue (timer expiration) and there are IP packets in the queue.
\end{itemize}

\subsection{Packets sending and scheduling principle}
\label{subsec::n2_scheduling}

\begin{figure}[h]
    \begin{center}
	\includegraphics[width=1\linewidth]{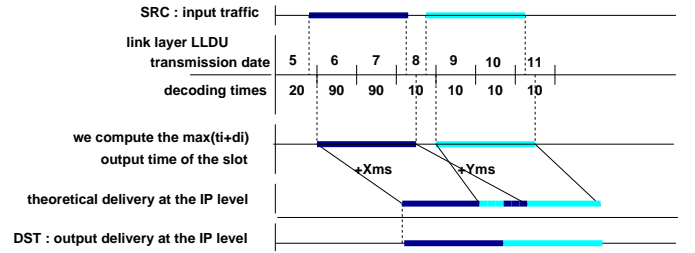}
	\caption{2 packets sharing channel and $ns$-2}
	\label{fig::two_packet_channel}
    \end{center}
\end{figure}

\begin{figure}[h]
    \begin{center}
	\includegraphics[width=0.9\linewidth]{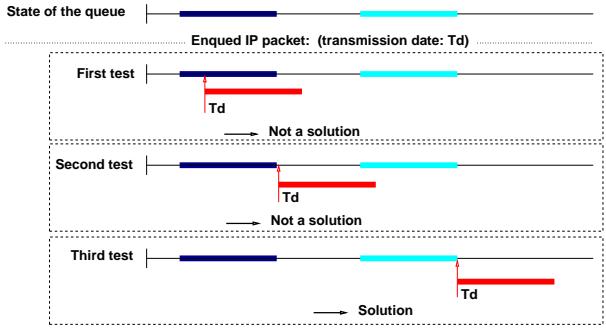}
	\caption{Adaptation of the transmission date of the IP packet}
	\label{fig::new_scheduling}
    \end{center}
\end{figure}

Figure~\ref{fig::two_packet_channel} illustrates the problem occurring when LLDU reliability schemes overlap the IP packets in terms of channel occupancy. In Figure~\ref{fig::new_scheduling} we detail the different cases we had to consider since $ns$-2 prevents one node from sending two packets at the same time.

If one of the LLDU is erased, the whole IP packet is dropped. Moreover the date of this event is linked to the reliability scheme introduced at the link layer. Indeed, the computed transmission date becomes the drop date. We also consider that a dropped packet still uses the channel for its transmission and has to be considered in the scheduling detailed in Figure~\ref{fig::new_scheduling}.

\section{Use case example}
\label{sec::illustration}

In this section, we show one example of the use-cases that CLIFT enables to study.
\subsection{Simulation definition}
\label{subsec::illustration_definition}

\subsubsection{Network and objectives}
We study the impact of retransmissions at the link layer on the performance of the transport protocol in a high bandwidth-delay product context. We consider a link between a satellite and a mobile receiver. In the following we present the characteristics of the channel, the link layer and the transport layer.

\subsubsection{Physical layer characteristics}
The physical layer trace corresponds to a mobile receiver moving at 60\,km per hour. The simulation lasts 400\,seconds. The size of the physical layer data unit is 33\,bytes and the capacity 2,3\,Mbps. We consider an interleaving at the physical layer of 35,5\,ms and and coding ratio of 1/3, waveform suitable for LTE uplink signals. In accordance with the phenomena described in~\cite{lms_state}, the data obtained introduces realistic signal-to-noise ratio variations (and burst erasures), modulations, multiplexing or frequency. The physical layer traces have been provided by CNES.

\subsubsection{Link layer characteristics}
In this example we study the impact of retransmissions at the link layer level on the performance of transport protocols. Therefore within the different reliability schemes introduced (detailed in ~\ref{subsec::link_layer_schemes}), we focus on ARQ and HARQ of type II. After an analysis of the channel state, for HARQ of type II, we set $N_D=10$ and $N_R=2~or~5$. The LLDU packet size is set to 33\,bytes.

\subsubsection{Transport layer characteristics}
The transport protocol used is TCP NewReno, implemented in $ns$-2. The IP packet size is set to 500 bytes. At the receiver side, we introduce a SACK mechanism. We aim to show the impact of the retransmissions at the link layer on the congestion window size. In order to better assess the impact of congestion window reduction, we limit the congestion window to $64$ IP packets. The ability of a transport protocol to reach the optimal congestion window will be studied as a future work.

\subsection{Results and interpretation}
\label{subsec::illustration_interpretation}
\begin{figure}[h]
    \begin{center}
	\includegraphics[width=1\linewidth]{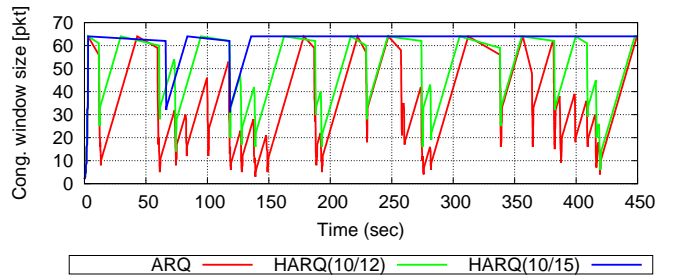}
	\caption{Congestion window evolutions}
	\label{fig::cwnd_arq_harq}
    \end{center}
\end{figure}

In this section, before interpreting the results, we collect the different metrics obtained during this simulation, in terms of: 
\begin{itemize}
\item used resources, goodput, mean coding ratio (MCR), delay and retransmission distribution; information is gathered in Table~\ref{tab::out_of_simu};
\item congestion window evolution and packet transmission: we plot in Figure~\ref{fig::cwnd_arq_harq} the evolution of the congestion window depending on the link layer reliability schemes introduced.
\end{itemize}

\begin{table}[h]
\caption{Measures obtained after a simulation}
\label{tab::out_of_simu}
\begin{center}
\begin{tabular}[h]{|c|c|c|c|c|}
\hline
\multicolumn{2}{|c|}{Metrics} & ARQ & \multicolumn{2}{|c|}{HARQ} \\
\cline{4-5}
\multicolumn{2}{|c|}{} & & 10/12 & 10/15 \\
\hline
\multicolumn{2}{|c|}{\% of the bandwidth used} & 13\% & 17\% & 23\% \\
\hline
\multicolumn{2}{|c|}{Goodput (kbps)} & 305 & 351 & 390 \\
\hline
\multicolumn{2}{|c|}{MCR ($\frac{useful~data}{sent~data}$)} & 95\% & 80\% & 65\% \\
\hline
 & minimum & 287 & 287 & 288 \\
\cline{2-5}
delay (ms) & mean & 288 & 288 & 289 \\
\cline{2-5}
 & maximum & 383 & 341 & 327 \\
\hline
Retransmission & 0 & 99\% &  98\% &  99\% \\
\cline{2-5}
number & 1 & 0,7\% & 1,4\% & 0,6\% \\
\cline{2-5}
(link layer) & 2 & 0,04\% & 0,03\% & 0,01\% \\
\cline{2-5}
 & 3 & 0,0007\% & 0\% & 0\%\\
\hline
Retransmission & 0 & 98\% & 99\% & 99\% \\
\cline{2-5}
number & 1 & 1,5\% & 0,6\% & 0,7\% \\
\cline{2-5}
(transport layer) & 2 & 0,1\% & 0,1\% & 0\% \\
\hline
\end{tabular}
\end{center}
\end{table}

With the data gathered in Table~\ref{tab::out_of_simu}, we can see that HARQ\,(10/15) has the best performance in terms of goodput and delay. Thereby, as more repair packets are sent, more bandwidth is used for this only application. Moreover, we can notice that even if we do not optimise the value of $N_D$ nor the ratio between $N_D$ and $N_D+N_R$ HARQ of type II enables the transmission of more data than an SR-ARQ reliability scheme, but use more bandwidth. 

We can see that there are more retransmissions at the transport layer with a SR-ARQ mechanism at the link layer. In consequence, we also see that the congestion window is reduced more often. Indeed, this can be explained by the fact that, while this mechanism enables the recovery of data, the IP packet is received after an additional delay. As the delayed IP packet is not acknowledged, the transport protocol assumes that if has been lost. When the congestion window is large, the delayed acknowledgements introduce spurious retransmissions and might greatly deteriorate the transmission of data as there is a reduction of the size of congestion window.

Through this example, we illustrated that, on a channel with a high erasure probability, with realistic parameters and bursty aspects, an SR-ARQ mechanism can introduce an important number of spurious retransmissions and reductions of the congestion window size: as the retransmissions modify the scheduling of the IP packets, the non-acknowledgement of some IP packets greatly deteriorate the performance of TCP NewReno protocol. As a future work we aim to study and observe the impact of retransmissions at the link layer level on the performance of the most recent transport protocols implemented in $ns$-2.

We illustrated here that retransmissions at the link layer can greatly deteriorate the performance of a loss-based transport protocol. As an HARQ-II mechanism first sends a FEC block, it improves the performance in the simulation context. We considered a maximal congestion window of 64 packets. It would be interesting to study the impact of the bandwidth reduction due to the transmission of these repair packets. When the capacity of the link is reached, a trade-off has to be found between reducing the congestion window (with SR-ARQ) and reducing the available bandwidth (HARQ-II).




\section{Conclusion}
\label{sec::conclusion}


We have presented a software that enables cross-layer studies between transport and MAC/PHY layers. We have developed the Cross Layer InFormation Tool (CLIFT), a simulator based on $ns$-2 that takes into account physical or link layer traces to schedule the transmission of transport layer packets. 

Our software can accodomate several networks architectures while taking into account most recent transport protocols. The originality of our tool consists in taking into account realistic sets of physical layer parameters (coding ratio, modulation, waveform). An important variety of existing tools can provide traces loaded in CLIFT, as they can be measured or simulated. 
In this article, we focused on satellite links. However, CLIFT can take into account any physical layer traces (Wi-Fi, wired or satellite links) in the context where cross-layer studies are of interest. 

Our tool can contribute to study current problems introduced by the augmentation of satellite links in recent networks. As future work,  we aim to perform a realistic study of transport layer protocols performance in the challenging context of satellite communications. Modeling the physical channel is complex and may require approximations, the use of CLIFT will reduce the potential erroneous performance evaluations. 

Finally, we also expect to assess the impact of various link layer schemes on the most recent transport protocols. We strongly believe that, in the context of Performance Enhancing Proxy\cite{pep_1,pep_2} or in aeronautical communications, considering different link layer reliability schemes on the satellite link will provide interesting measurements to evaluate the transport protocols performance. 

\section*{Acknowledgement}
This work has been funded by CNES. The authors wish to thank Thales Alenia Space for technical support and Olivier Mehani for help and comments.

\end{document}